# Drude Conductivity of Dirac Fermions in Graphene

Jason Horng[1], Chi-Fan Chen[1], Baisong Geng[1], Caglar Girit[1], Yuanbo Zhang[2], Zhao Hao[3,4], Hans A. Bechtel[3], Michael Martin[3], Alex Zettl[1,5], Michael F. Crommie[1,5], Y. Ron Shen[1,5] and Feng Wang*[1,5]

[1] Department of Physics, University of California at Berkeley, Berkeley, CA 94720, USA

[2] Department of Physics, Fudan University, Shanghai, 200433, China

[3] Advanced Light Source Division, Lawrence Berkeley National Laboratory, Berkeley, CA 94720, USA

[4] Earth Sciences Division, Lawrence Berkeley National Laboratory, Berkeley, CA 94720, USA

[5] Materials Science Division, Lawrence Berkeley National Laboratory, Berkeley, CA 94720, USA

* To whom correspondence should be addressed. Email: fengwang76@berkeley.edu

**Abstract:**

Electrons moving in graphene behave as massless Dirac fermions, and they exhibit fascinating low-frequency electrical transport phenomena. Their dynamic response, however, is little known at frequencies above one terahertz (THz). Such knowledge is important not only for a deeper understanding of the Dirac electron quantum transport, but also for graphene applications in ultrahigh speed THz electronics and IR optoelectronics. In this paper, we report the first measurement of high-frequency conductivity of graphene from THz to mid-IR at different carrier concentrations. The conductivity exhibits Drude-like frequency dependence and increases dramatically at THz frequencies, but its absolute strength is substantially lower than theoretical predictions. This anomalous reduction of free electron oscillator strength is corroborated by corresponding changes in graphene interband transitions, as required by the sum rule. Our surprising observation indicates that many-body effects and Dirac fermion-impurity interactions beyond current transport theories are important for Dirac fermion electrical response in graphene.

Graphene provides a unique material system to study Dirac fermion physics in two dimensions. Researchers have demonstrated in graphene exotic Dirac fermion phenomena ranging from anomalous quantum Hall effects[1,2] to Klein tunneling[3] in low frequency (DC) electrical transport. They also observed an optical conductance defined by the fine structure constant[4,5] and gate-tunable infrared absorption in Dirac fermion interband transitions[6,7]. Situated between low-frequency electrical transport and interband optical excitation is the spectral range dominated by free carrier intraband transitions, studies of which can provide new insight into both Dirac fermion transport and interband transitions. This free carrier dynamics response is also expected to play a key role in future development of ultrahigh-speed electronics at THz frequencies and THz-to-mid-IR optoelectronic devices[8,9]. However, despite its importance, experimental study of free carrier response of graphene in the THz to mid-IR spectral range has been lacking.

Quantum theories of electron dynamic response in graphene have been developed by several groups [10-15]. They predict that, within the framework of Boltzmann transport theories, high frequency conductivity of monolayer graphene has a Drude form, $\sigma(\omega) = \frac{iD}{\pi(\omega + i\Gamma)}$, where $\omega$ is the frequency and $\Gamma$ the scattering rate. The prefactor $D$, known as the Drude weight, has the value $D = \frac{v_F e^2}{\hbar}\sqrt{\pi|n|}$ when electron-electron interactions are neglected, with $v_F = 1.1 \times 10^6 \, m/s$ being the Fermi velocity[2] and $|n|$ the carrier density. This is distinctly different from classical materials where $D = \frac{\pi n \cdot e^2}{m^*}$. The predictions on graphene Dirac fermion dynamics have never been directly tested experimentally. In fact, it was suggested from measurements of interband transitions [6] that D could deviate from the predicted value. There are also theoretical studies showing that electron-impurity interactions could result in carrier dynamics in graphene beyond the Boltzmann description [16,17], and that inclusion of electron-electron interactions could dramatically reduce the Drude weight $D$ [18].

Here, we report spectroscopic measurements of graphene conductivity from terahertz to mid-IR for different electron and hole concentrations. We confirm that the conductivity from free-carrier response indeed has a Drude-like frequency dependence. In addition, we have been able to determine the Drude weight $D$ and scattering rate $\Gamma$ of doped graphene directly and separately for the first time. We observe an electron-hole asymmetry for $D$, and find that values of $D$ is substantially lower than the theoretical prediction $D = \frac{v_F e^2}{\hbar}\sqrt{\pi|n|}$ [10-12]. It indicates that a better understanding of electric transport and Drude conductivity need to include electron-electron and electron-impurity interactions beyond the current theories for carrier dynamics. This reduction of Drude weight (arising from free carrier intraband transitions) is connected to corresponding changes in interband transitions (6), and we demonstrate that the sum rule requiring the integrated intraband (Drude) conductivity and interband conductivity to be a constant is well obeyed.

In our study we used Fourier transform infrared spectroscopy (FTIR) to measure the transmission spectra of graphene samples over the range from 30 to 6000 cm$^{-1}$ and deduce the frequency-dependent conductivity from the spectra. Previously, infrared spectroscopy to probe graphene monolayer was limited to >1000 cm$^{-1}$ (wavelength < 10 $\mu m$) because of the limited size of exfoliated graphene [4,6,7,19]. For spectroscopy at longer wavelength up to ~300 $\mu m$, we need large sample size. Here we used large-area graphene grown by chemical vapor deposition (CVD). (Fig. 1a) (see methods)

Fig. 1b shows the DC conductivity as a function of the gate voltage in a typical graphene sample. The minimum conductance occurs at 14 V, which defines the charge neutral point (CNP) of the sample. Carrier density in graphene is related to the gate voltage by $n = 7.5 \times 10^{10} (V_{cnp} - V_g) \, cm^{-2} V^{-1}$ [2], where negative (positive) density corresponds to electron (hole) doping. The sample had an initial hole doping of $1.05 \times 10^{12}$ cm$^{-2}$ at $V_g = 0$, and a hole mobility around 2700 cm$^2$/V.s. As seen from Fig. 1b, the DC conductivity for hole doping is reasonably linear with $|n|$, and there is a large asymmetry between electron and hole conductivity at the same $|n|$. In previous studies, conclusions have frequently been made on carrier scattering from such DC conductivity data based on $\sigma_{DC} = D/\pi\Gamma$ and the assumption that $D = \frac{v_F e^2}{\hbar} \sqrt{\pi |n|}$ holds exactly as theory predicts[21-23]. However, the validity of this approach has never been tested.

An independent determination of $D$ and $\Gamma$ can be achieved through AC conductivity measurements using IR spectroscopy. Fig. 1c shows a difference IR absorption spectrum of hole-doped graphene ($V_g$ = -70V) in reference to absorption at the CNP ($V_g$ = 14V). This difference absorption spectrum has two characteristic features: a large absorption increase at lower wavenumbers, which reaches over 15% at terahertz frequencies for a graphene monolayer, and an absorption reduction over a broad range of higher wavenumbers. These features can be understood qualitatively from the gate-induced changes in intraband and interband electronic transitions in graphene due to carrier doping (Fig. 1d). Free carrier conductivity (from intraband transitions) increases dramatically with the hole doping. It peaks at zero frequency and gives rise to a substantial absorption increase at low wavenumbers. At the same time, however, the interband transitions up to energy $2E_F$ (arrows in Fig. 1d) become forbidden due to empty initial states, leading to a reduction of absorption in the broad spectral range below $2E_F$ (dashed line in Fig. 1c).

The gate-induced change of AC conductivity (referred to CNP value), $\Delta\sigma = \sigma - \sigma_{cnp}$, can be obtained readily from the difference IR absorption spectra (see supplementary information). Fig. 2a and Fig. 2b show the real part of the AC conductivity change $\Delta\sigma'$ in the low-wavenumber range (<450 cm$^{-1}$) for different electron and hole concentrations. In this spectral range free carrier response (i.e., an increase in conductivity with carrier density) dominates, providing direct information on Dirac fermion electrical transport. The complete set of $\Delta\sigma'$ spectra in Figs. 2a and 2b can be fit (dashed line) by $\Delta\sigma(\omega) = iD/\pi(\omega + i\Gamma) - iD_{cnp}/\pi(\omega + i\Gamma_{cnp})$ using the Drude form for $\sigma$ and $\sigma_{cnp}$. A finite charge neutral point response ($D_{cnp}$ and $\Gamma_{cnp}$) accounts for inhomogeneous electron and hole puddles present in graphene. In the fittings, we set $\Gamma_{cnp}$ equal to that when graphene is weakly doped. The value of $D_{cnp}$ is set so that $D_{CNP}/\pi\Gamma_{CNP}$ gives the DC conductance at the charge neutral point. From the Drude fits, we obtain directly the scattering rate $\Gamma$ and Drude weight $D$ as a function of electron and hole concentrations, shown as symbols in Fig. 2c and Fig. 2d, respectively.

From the values of $D$ and $\Gamma$ extracted from the IR spectroscopy, we can obtain $\sigma_{DC}(=D/\pi\Gamma)$ of graphene at different carrier concentrations (red dots in Fig. 2e). It agrees nicely with the directly measured DC conductivity in our sample for all gate voltages (black dots). This agreement provides a consistency check of the Drude description.

We now compare our experimentally determined $\Gamma$ and D to theory predictions. Figure 2c shows that $\Gamma$ is roughly a constant at different hole concentrations, while it is higher in value and increases with the carrier concentration for electron doping. This electron/hole concentration dependence of $\Gamma$ is not predicted by theory, and cannot be described by pure unitary scattering or charge impurity scattering of Dirac fermions[10-12,24]. A combination of different scattering mechanisms seems to be necessary to explain the observation.

While theoretical prediction on $\Gamma$ may vary depending on the scattering mechanisms assumed, the same theoretical value for the Drude weight $D = \frac{v_F e^2}{\hbar}\sqrt{\pi n}$ [10,12,15,25] is always obtained. Comparison of theory (blue line) and experiment (black symbols) in Fig. 2d, however, shows an appreciable difference. In the strongly doped region ($|V_g-V_{cnp}|>20$ V) where the Boltzmann theory is supposed to be most reliable, the measured D is lower than the theoretical value by 20-45%. (The experimental uncertainty is less than 10%.). An electron and hole asymmetry is also present in the measured D. The experimentally observed reduction of Drude weight is quite unexpected and it suggests that the prevailing graphene transport theory is not complete.

The general sum rule for oscillator strengths in solids state that the integrated conductivity over all frequencies is a constant, i.e. $\int_0^\infty \Delta\sigma'(\omega)d\omega = 0$. In the case of non-interacting free carriers described by prevailing graphene theories, the integrated intraband absorption increases by $\int \Delta\sigma'_{intra} d\omega = \frac{D}{2} = \frac{v_F e^2}{2\hbar}\sqrt{\pi |n|}$ upon carrier doping[10-12].

This increase is compensated exactly by an integrated interband absorption $\int \Delta\sigma'_{inter}\, d\omega = \sigma_0 \cdot (2E_F/\hbar)$ with $E_F = \hbar v_F \sqrt{\pi|n|}$ and $\sigma_0 = e^2/4\hbar$ (Fig. 1d.) [4,5,26-28]. In our case, the sum rule is still valid, and the observed anomalous reduction of Drude weight in intra-band transitions should be accompanied by a corresponding change in interband transitions at large wavenumbers. Fig. 3a and 3b show the optical conductivity difference (referred to CNP) at different hole and electron concentrations in the spectral range 600-6000 cm$^{-1}$. A gated-induced decrease in optical conductivity is clearly present from zero to a cut-off frequency of $\dfrac{2E_F}{\hbar}$ that increases with carrier doping, as the theory predicts. However, the observed interband optical conductivity decrease in doped graphene is appreciably less than $\sigma_0$ (dashed lines in Fig. 3a and 3b), and the deviation is larger for electron doping compared with hole doping. This reduction in gate-induced absorption decrease, as well as electron-hole asymmetry in interband transitions, matches well with the reduction in gate-induced absorption increase and electron-hole asymmetry in intraband Drude weight. To be more quantitative, we plot in Fig. 3c the absolute values of integrated conductivity change for interband transitions (black symbol) and intraband transitions (red symbol, which is proportional to the Drude weight change) at different carrier doping. The two agree almost perfectly, just as the sum rule requires. A reduction of gate-induced interband absorption has also been observed previously for mechanically exfoliated graphene[6]. Based on the sum rule that we established experimentally, it indicates that the anomalous Drude weight reduction is general for both exfoliated and CVD samples.

That the Drude weight D is significantly smaller than the predicted value of $D = \frac{v_F e^2}{\hbar}\sqrt{\pi n}$, seen directly from the measurements of the Drude conductivity and confirmed by the interband absorption spectra, shows that modifications on the prevailing transport theory[10-12] is needed to describe Dirac fermions in graphene. Inclusion of electron-electron interactions and improved treatment of electron-impurity interactions may be necessary. It has already been shown in Ref. [18] that, unlike conventional massive fermions, the Drude weight of massless Dirac fermions can be strongly reduced by electron-electron interactions due to pseudo-spin physics, although this theory predicts a much larger reduction of the Drude weight (over 80%) than what we have observed. In addition, the electron-impurity interactions may play a role and contribute to the observed electron-hole asymmetry[17]. It is also possible that reduction of D is a consequence of electron or hole localization, which decreases the effective 'free' electron or hole concentration. This localization, however, clearly goes beyond impurity scattering of Dirac fermions considered in current quantum transport theory[10,12,25]. In any case, a complete theory must both explain the reduction of Drude weight and provide a consistent description of the absorption changes in interband transitions.

Our infrared spectroscopy shows that the frequency-dependent response of Dirac fermion in graphene can be described by the Drude form, which becomes very strong at longer wavelengths. In the THz range ($\lambda \sim 300 \mu m$), a gated graphene monolayer can absorb over 15% of the incident radiation, suggesting that graphene can potentially be a useful new THz material. More importantly, the independent determination of Drude weight and scattering rate for the free carriers in graphene has allowed us to discover that the free carrier Drude weight is significantly smaller than the widely-accepted theoretical predictions, but the sum rule for integrated intraband and interband oscillator strength is strictly obeyed.

**Methods**

Following the procedure in Ref. [20], we grew graphene on copper films using $CH_4$ as the feed gas, which was then transferred with PMMA support to a $Si/SiO_2$ wafer after wet-etching to remove the copper film by $FeCl_3$. After dissolving the PMMA in acetone solution, high-quality graphene of *cm* size on the $Si/SiO_2$ wafer was obtained. (Fig.1a) Raman spectra show that the sample was indeed monolayer graphene. Subsequently, Au/Ti electrodes (thickness ~ 50 nm) were deposited in vacuum through stencil masks onto the graphene sample for electrical measurements. The doped Si substrate (p-type, resistivity ~ 10 $\Omega \cdot cm$) underneath the 290 nm $SiO_2$ served as a back gate, which allowed us to tune the charge carrier density in the graphene sample. All optical and electrical measurements were performed in vacuum (~ 0.1 mTorr) at 100 K.

**Acknowledgements**

This work was supported by the U.S. Department of Energy, Laboratory Directed Research and Development Program of Lawrence Berkeley National Laboratory under Contract No. DE-AC02-05CH11231, the Office of Basic Energy Sciences under contract No. DE-AC03-76SF0098 (Materials Science Division) and contract No. DE-AC02-05CH11231 (Advanced Light Source), and ONR MURI award N00014-09-1-1066.

**Author Contribution**

F.W. designed the experiment. J.H, C.F.C. Z.H, H.A.B. carried out optical measurements, C.F.C. B.G., C.G. and Y.Z. contributed to the sample growth and fabrication, J.H. and F.W performed the data analysis. All authors discussed the results and wrote the paper together.

**Competing interests statement**

The authors declare no competing financial interests.

Figures:

**Figure 1 Properties of CVD-grown graphene device. a**, Optical microscope image of CVD-grown graphene transferred to a SiO$_2$/Si substrate. **b,** Graphene DC conductivity as a function of gate voltage. The conductivity minimum at V$_g$ = 14V defines the charge neutral point (CNP). **c,** Gate-induced change of IR transmittance –ΔT/T through graphene at V$_g$ = -70 V (hole doped) compared to transmittance at the CNP. The spectrum shows an increase of free-carrier absorption at low wavenumbers and a reduction of interband absorption at higher wavenumbers. **d,** An illustration of intraband (i.e. free-carrier absorption, purple arrow) and interband (blue arrow) transitions in hole-doped graphene. Intraband absorption increase with carrier doping, while interband transitions up to 2E$_F$ become forbidden due to empty initial states, as observed in c.

**Figure 2 Free-carrier responses in graphene. a,b,** Gate-induced change of AC conductivity $\Delta\sigma'$ in hole- and electron-doped graphene, respectively, for 30 cm$^{-1}$< ω <450 cm$^{-1}$ (solid lines). The frequency dependence of AC conductivity can be fit by the Drude model (dashed lines). **c**, The scattering rate Γ and ,**d**, Drude weight *D* at different electron and hole concentrations. Surprisingly, the Drude weight is substantially lower than theoretical predictions based on the Boltzmann transport theory (blue line in **d**) and shows an electron-hole asymmetry. **e**, DC conductivity in the form of D/πΓ obtained from the Drude fit of terahertz/far-IR measurements (red dots) agrees well with the values from direct DC transport measurements (black dots).

**Figure 3 Interband transitions in graphene. a,b,** Gate-induced change of interband AC conductivity $\Delta\sigma'$ in hole- and electron-doped graphene, respectively, over 600 $cm^{-1}$ < ω < 6000 $cm^{-1}$. The AC conductivity decreases due to empty initial states (hole doping) or filled final states (electron doping) for interband transitions below 2$E_F$. (The sharp features around 1200 $cm^{-1}$ are extrinsic and arise from $SiO_2$ phonon resonances of the substrate). Observed interband conductivity decrease is appreciably less than the theoretical predicted value of $\pi e^2/2h$ (horizontal dashed line), and shows electron-hole asymmetry. **c,** Upon charge carrier doping, integrated conductivity change has an increase from intraband transitions (red symbols) that equals the decrease from interband transitions (black symbols). The two values agree quantitatively at all gate voltages, as required by the oscillator strength sum rule.

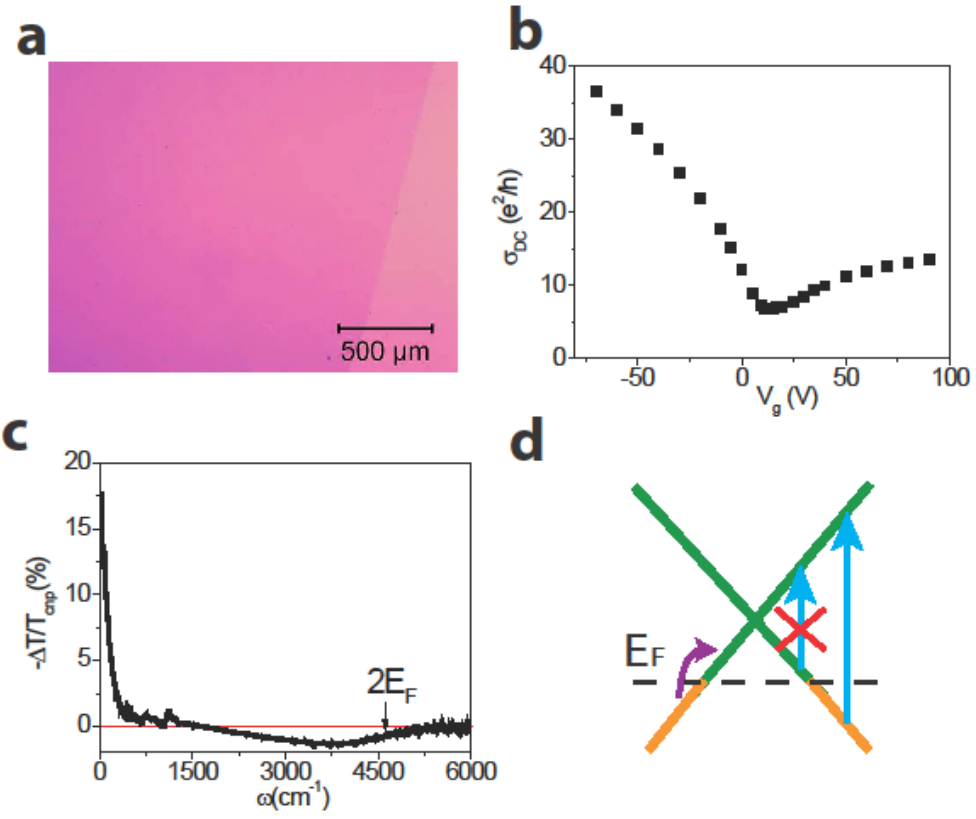

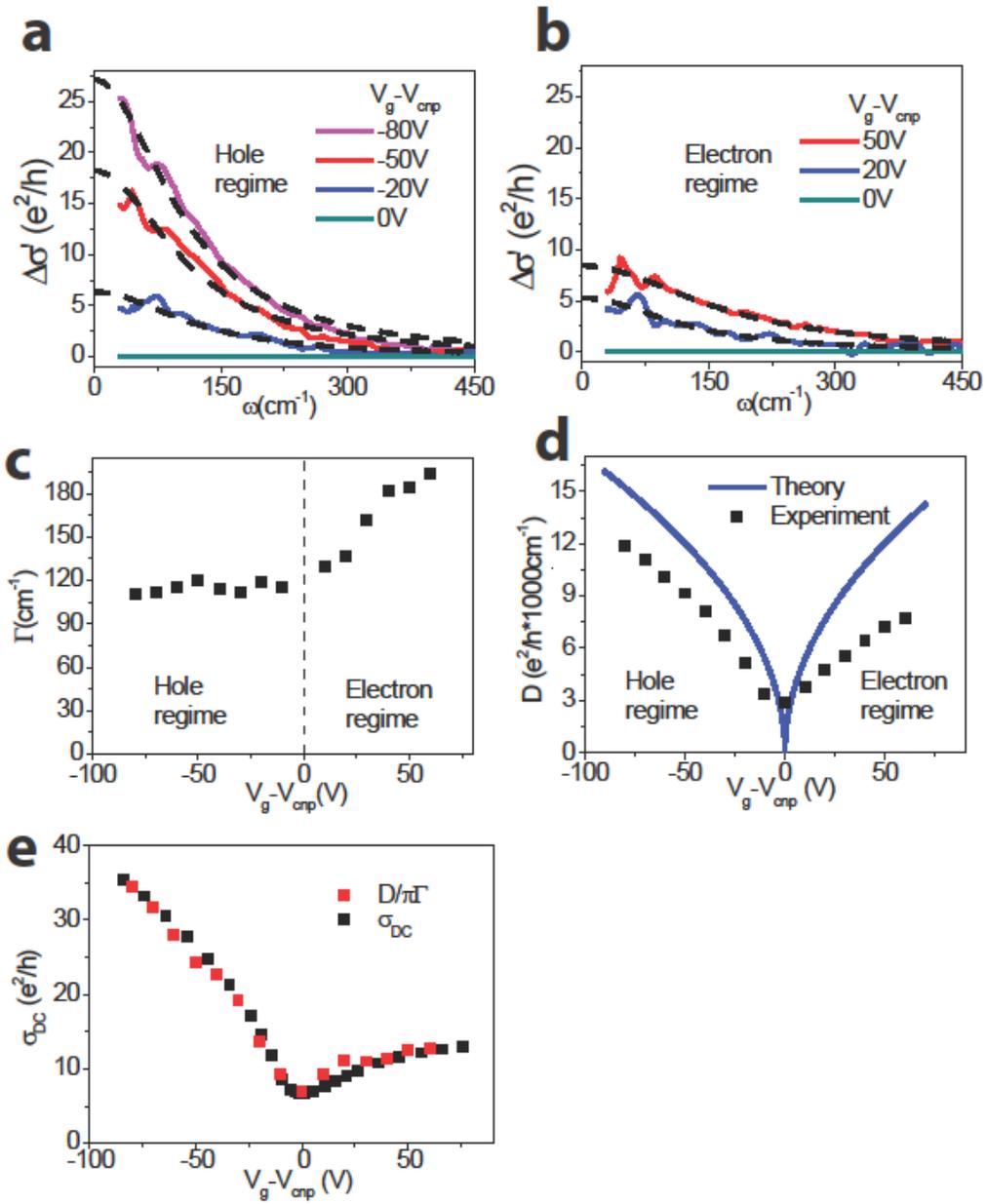

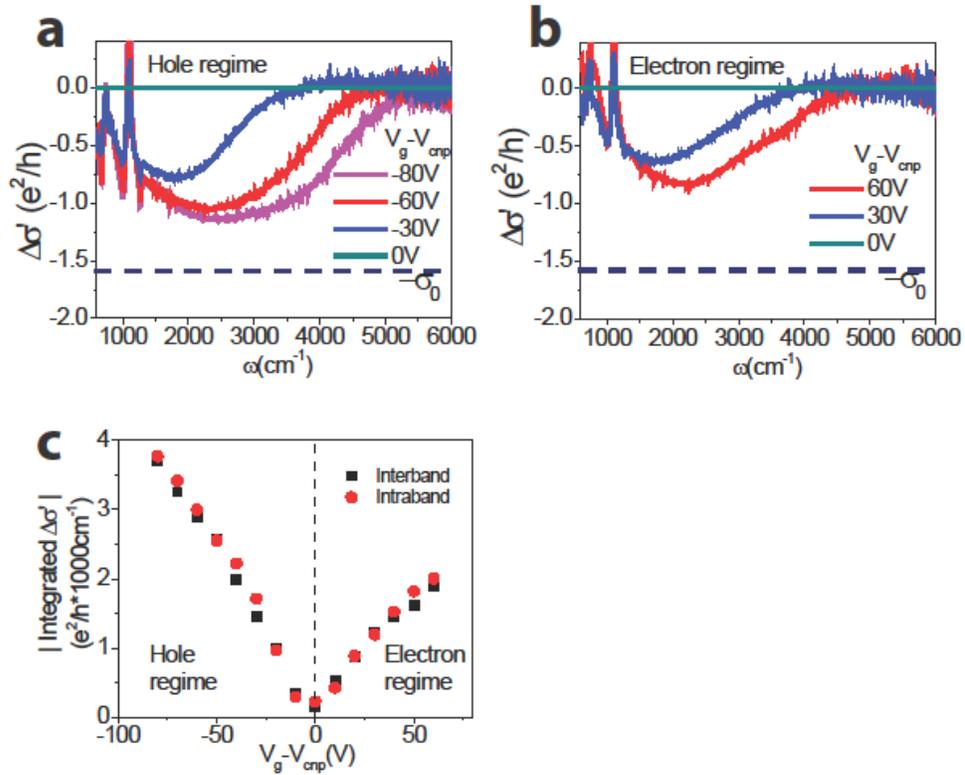

Supplementary Information:

The frequency-dependent conductivity change of graphene $\Delta\sigma$ at different gate voltages is obtained from the measured transmission spectra using a perturbation treatment, because a monolayer graphene only absorbs a small fraction of the light. Within this approximation, the complex AC conductivity change is related to the difference transmission spectra by

$$\frac{-\Delta T}{T}(\omega) = \frac{4\pi}{c}\text{Re}(\Delta\sigma \times L) \qquad (\text{S\_Eq. 1})$$

where $-\Delta T/T$ is the normalized transmission difference, L the local field factor, and $\Delta\sigma$ the gate-induced conductivity change in graphene.

For suspended graphene, the local field factor is one and we obtain the well know form of $\frac{-\Delta T}{T}(\omega) = \frac{4\pi}{c}\cdot\Delta\sigma'$ (S1). In our device, the graphene is sitting on a $SiO_2$/Si substrate and the local field factor L is uniquely defined by the refractive index of Si and $SiO_2$, as well as the thickness of $SiO_2$, and it can be determined accurately without any adjusting parameters. Specifically, L has the form

$$L = 1 + r_{ag} + t_{ag}r_{gs}t_{ga}\frac{e^{2ikn_g d}}{1 - e^{2ikn_g d}r_{gs}r_{ga}} \qquad (\text{S\_Eq. 2})$$

where $r_{ag}$, $r_{gs}$ and $r_{ga}$ are the Fresnel reflection coefficients at air-silica, silica-silicon and silica-air interferences, $t_{ag}$ and $t_{ga}$ are the corresponding transmission coefficients, $n_g$ and $d$ are complex reflective index and thickness of the $SiO_2$ layer, respectively, and $k$ is the wavevector of the incident light.

The real and imaginary part of AC conductivity, $\sigma'$ and $\sigma''$, is connected by Kramers-Kronig (K-K) relation

$$\sigma''(\omega) = \frac{-2\omega}{\pi} \int_0^\infty \frac{\sigma'(\bar{\omega})d\bar{\omega}}{\bar{\omega}^2 - \omega^2} \qquad (S\_Eq.\ 3)$$

Because the K-K relation holds true for all the gate voltages, the gate-induced conductivity change $\Delta\sigma = \Delta\sigma' + i\Delta\sigma''$ also satisfies the same K-K relation. Making use of the S_Eq. 1 and the K-K relation, we obtain both the real and imaginary part of the gate-induced AC conductivity from infrared transmission spectra.

The gate-induced infrared absorption from the p-doped Si substrate is negligible. Real part of hole AC conductivity in silicon is described by $\sigma' = \frac{n \cdot e^2}{m_h} \frac{\Gamma}{\Gamma^2 + \omega^2}$. For the same carrier density, it is more than one order of magnitude smaller than graphene conductivity across the experimental spectral range (30 cm$^{-1}$ < ω < 6000 cm$^{-1}$) due to the large hole effective mass ($m_h = 0.36\ m_0$) and small scattering rate ($\Gamma \sim 4$ cm$^{-1}$ at 100 K) (S2).